\documentclass[%
reprint,
 amsmath,amssymb,
 aps,
 prx,
]{revtex4-2}

\usepackage{graphicx}
\usepackage{dcolumn}
\usepackage{bm}
\usepackage{xr}
\usepackage{notes2bib}
\usepackage{subfigure}
\usepackage{appendix}
\usepackage{quantikz}
\usepackage{enumerate}
\usepackage{enumitem}
\usepackage{hyperref}

\usepackage{accents}
\newcommand{\dbtilde}[1]{\accentset{\approx}{#1}}
\DeclareMathOperator{\Tr}{Tr}

\newcommand\id{1\kern-0.25em\text{l}}

\begin{document}


\title{Dimensionality reduction for closed-loop quantum gate calibration}

\author{Emma Berger$^{1,2}$}
\email{These authors contributed equally during internships at IBM Quantum.}
\author{Vivek Maurya$^{1,3,4}$}
\email{These authors contributed equally during internships at IBM Quantum.}
\author{Z.~M.~McIntyre$^{1,5}$}
\email{These authors contributed equally during internships at IBM Quantum.}
\author{Ken Xuan Wei$^1$}
\author{Holger Haas$^1$}
\author{Daniel Puzzuoli$^6$}
\email{Corresponding author: Daniel.Puzzuoli1@ibm.com}
\affiliation{${}^1$IBM Quantum, T.J. Watson Research Center, Yorktown Heights, NY, 10598, USA}
\affiliation{${}^2$Department of Physics, University of California, Berkeley, Berkeley, CA 94720, USA}
\affiliation{${}^3$Center for Quantum Information Science and Technology, University of Southern California, Los Angeles, CA, 90089, USA}
\affiliation{${}^4$Department of Physics \& Astronomy, University of Southern California, Los Angeles, CA, 90089, USA}
\affiliation{${}^5$Department of Physics, McGill University, Montreal, QC, H3A 2T8, Canada}
\affiliation{${}^6$IBM Quantum, IBM Canada, Vancouver, BC, V6C 2T8, Canada}

\date{\today}

\begin{abstract}
Numerical gate design typically makes use of high-dimensional parameterizations enabling sophisticated, highly expressive control pulses. Developing efficient experimental calibration methods for such gates is a long-standing challenge in quantum control, as on-device calibration requires the optimization of noisy experimental data over high-dimensional parameter spaces. To improve the efficiency of calibrations, we present a systematic method based on SVD stacking for reducing the dimensionality of the parameter space traversed in gate calibration, starting from an arbitrary high-dimensional pulse representation. We use this approach to design and calibrate an $X_{\pi/2}$ gate robust against amplitude and detuning errors, as well as an $X_{\pi/2}$ gate robust against coherent errors due to a spectator qubit.

\end{abstract}

\maketitle

\section{Introduction}
High-fidelity quantum gates are a prerequisite for fault-tolerant gate-based quantum computing. In recent decades, numerical gate design using optimal control theory (OCT) has emerged as a promising field with the ability to automate the design of sophisticated control-pulse waveforms, starting from a Hamiltonian model~\cite{krotov1983iterative,khaneja2005optimal,doria2011optimal,rach2015dressing,machnes2018tunable,trowbridge2023direct,chadwick2023efficient}. Numerical gate design offers the potential for flexible design of control sequences in arbitrary systems, with the ability to instill properties into the control sequence that are not always straight forward to design analytically (e.g., robustness with respect to certain model parameters). Despite the potential benefits, however, these control pulses frequently perform poorly in experiment, due primarily to the necessity of starting off with a highly accurate Hamiltonian model and a perfect description of control signal distortions. For superconducting qubits in particular, this requirement is difficult to satisfy: Model inaccuracies may arise from system characterization errors~\cite{egger2014adaptive,cole2015hamiltonian,werninghaus2021leakage,wittler2021integrated}, the presence of two-level systems in the qubit's environment~\cite{klimov2018fluctuations,burnett2019decoherence,muller2019towards, lisenfeld2019electric,de2020two,wang2022towards}, and complicated electronic transfer functions~\cite{gustavsson2013improving, jerger2019situ,rol2020time}. Due to these issues, realizing the potential of numerically designed gates requires the development of reliable and scalable closed-loop experimental calibration methods.

A fundamental challenge in the calibration of numerically designed gates arises from the complicated, high-dimensional waveform parameterizations used in optimizations. These parameterizations allow the optimizer to flexibly explore the control space, but with the consequence that individual parameters have an unclear relationship with the resulting system evolution. This situation can be contrasted to the situation for gates derived from analytic solutions~\cite{paraoanu2006microwave,motzoi2009simple, rigetti2010fully,theis2018counteracting,magesan2020effective}, such as the Derivative Removal by Adiabatic Gate (DRAG)~\cite{motzoi2009simple}: For a DRAG pulse, the overall pulse amplitude sets the rotation angle of the qubit along a fixed axis in a way that is well understood, and calibration schemes are based on estimating this angle and adjusting the amplitude accordingly. For a numerically designed pulse with fully general two-axis control, changing any given pulse parameter could potentially induce changes to both the rotation angle and axis of rotation, making calibration challenging.

As a result of the black-box relationship between pulse parameters and system evolution, recent work has taken a more traditional optimization approach to calibration: formulating an objective function on experimental data and applying a general minimization algorithm on all pulse parameters. Algorithms that have previously been considered for this task include gradient-free optimization algorithms such as Nelder-Mead \cite{egger2014adaptive,nelder1965simplex}, as well as evolutionary algorithms such as the covariance-matrix adaptive evolution strategy (CMA-ES) \cite{werninghaus2021leakage,hansen2003reducing}. These approaches unfortunately suffer from long convergence times: Nelder-Mead, for instance, converges increasingly slowly as the number of parameters approaches ten~\cite{han2006effect}, and while CMA-ES can handle a larger ($\sim 50$) number of parameters, the convergence times are still prohibitive for the purposes of routine calibration. Other approaches include reinforcement learning~\cite{sivak2022model} and ``data-driven'' methods combining experimental data with model-based gradients~\cite{wu2018data}. However, as is generally the case with gradient-free optimizers, the successful application of these methods is heavily limited in practice by the high number of tunable pulse parameters typically used in numerical gate design.

For the benefits of numerical gate design to be fully realized in experiment, calibration procedures must therefore reconcile the disconnect between (i) fully general numerical gate design involving high-dimensional pulse parameterizations, and (ii) the non-trivial, black-box relationship between pulse parameters and system evolution. With the goal of addressing this disconnect, this work presents a systematic way of interfacing numerical gate design and calibration via a dimensionality reduction procedure designed to significantly reduce the number of pulse parameters that need to be calibrated experimentally. This method could in principle improve the efficiency of any calibration procedure by reducing the search space dimension, allowing numerical quantum control to be leveraged to its full potential in experiment. In what follows, we describe the dimensionality reduction procedure in Sec.~II. We then demonstrate the application of this approach in Sec.~III by designing and calibrating an $X_{\pi/2}$ gate robust against amplitude and frequency variations, as well as an $X_{\pi/2}$ gate robust against unwanted spectator interactions. Finally, we conclude by providing some outlook in Sec.~IV.

\begin{figure}
    \centering
    \includegraphics[width=\linewidth]{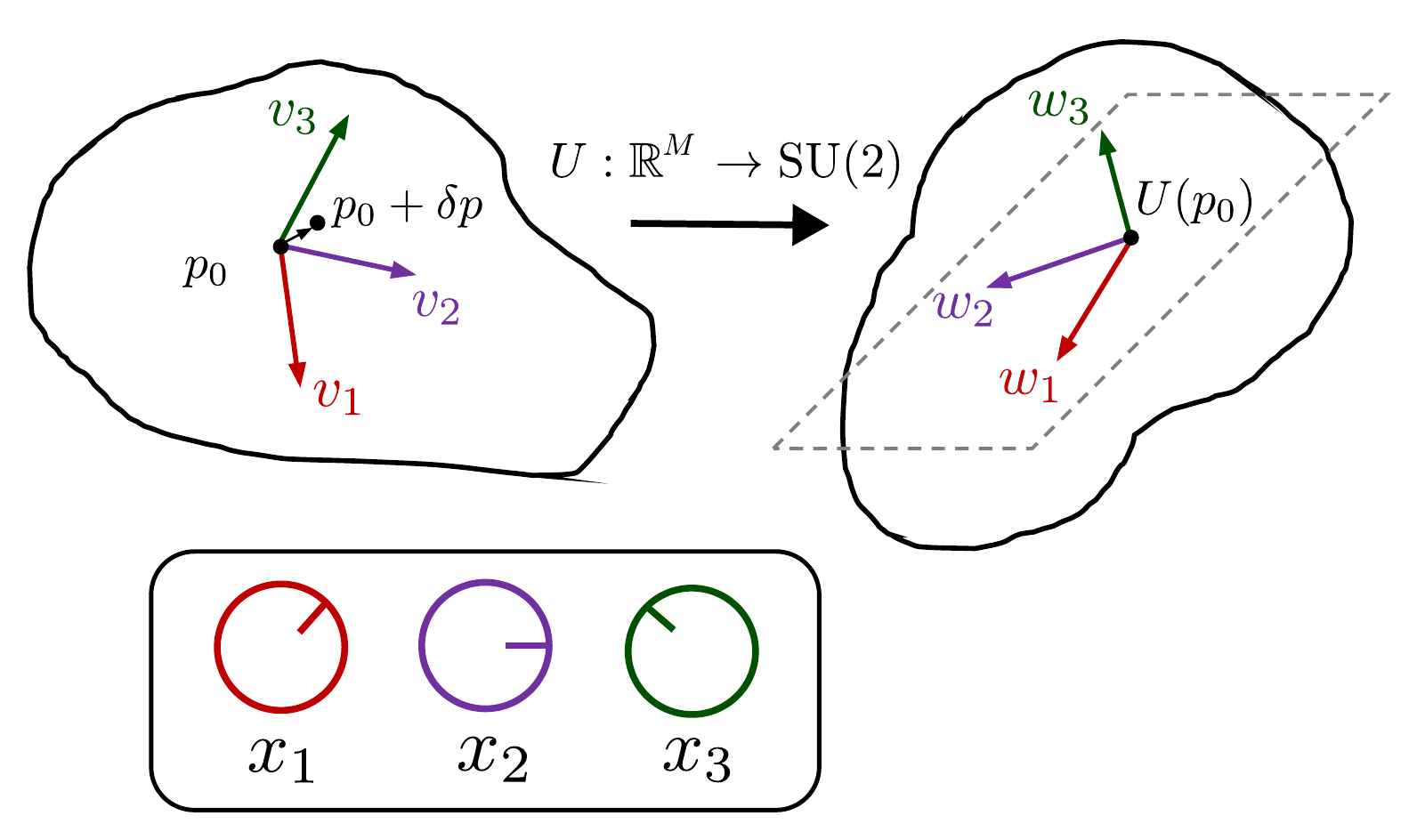}
    \caption{Dimensionality reduction procedure for closed-loop calibration of single-qubit gates: Numerical optimization provides a map $U : \mathbb{R}^M \rightarrow \mathrm{SU}(2)$ from the space of pulse parameters to the special unitary group $\mathrm{SU}(2)$. The SVD of this map, evaluated at the vector $p_0$ of optimized pulse parameters, permits the identification of a three-dimensional subspace of $\mathbb{R}^M$ (spanned by $v_1,v_2,v_3$) to which we may restrict the calibration vector $\delta p$  without restricting the corresponding change in $U(p_0+\delta p)$. This is because the basis vectors $w_1,w_2$ and $w_3$ corresponding (via the SVD) to $v_1,v_2$ and $v_3$ span the space tangent to SU(2) at $U(p_0)$. Calibration is performed by tuning $\bm{x}=(x_1,x_2,x_3)$ [cf Eq.~\eqref{delta-p}]. In practice, we use a generalization of this procedure (described in the text) that accounts for possible model-dependence of $p_0$. }
    \label{fig:dim_red}
\end{figure}

\section{Dimensionality reduction of the calibration search space}\label{sec:dimensionality_reduction}

In numerical gate design, the goal is to design a pulse envelope $s(t)$ so that evolution under some model Hamiltonian $H[s(t)]$ generates a target unitary (quantum gate). We may also require $s(t)$ to have additional properties, such as robustness to particular model variations: $H\mapsto H+\delta H$, where $\delta H$ may be treated in perturbation theory. In practice, the envelope $s(t)$ is parameterized by a vector $p$ of $M$ real numbers, which are numerically optimized according to an objective function encoding the desired pulse properties. While these pulse parameters could in principle consist of a discretization of the pulse envelope, as in Ref.~\cite{werninghaus2021high}, they correspond here to coefficients in a polynomial basis. Details of the parameterization are given in Appendix A.

In optimization problems involving hard-to-satisfy requirements like simultaneous robustness to multiple model variations, it is necessary for the dimension of the parameterization, $M$, to be sufficiently large to accommodate highly expressive pulse shapes during the initial gate design stage. Given a particular choice of pulse parameters $p_0 \in \mathbb{R}^M$ found via numerical optimization, the immediate subsequent challenge is to apply the resulting pulse in experiment. However, due to an almost inevitable mismatch between model inputs and experimental conditions (e.g.~the combined effects of the environment, imperfectly characterized electronic transfer functions, and interactions excluded from the model Hamiltonian), the gate fidelities initially achieved in experiment are typically much worse than predicted unless specific control-pulse-distortion characterization methods are employed \cite{Gustavsson2012,Hincks2015,Rose2018}. As a result, successful experimental application generally requires further calibration guided by experimental data~\cite{kelly2014optimal,werninghaus2021leakage}. However, for large $M$ ($M\gtrsim 10$), the process of tuning all $M$ parameters based on experimental data may be expensive~\cite{werninghaus2021leakage}. An approach for systematically reducing the dimensionality of the pulse thus provides a way to reduce calibration costs.

Starting from an initial point in parameter space $p_0$, the experimental calibration process consists of selecting modified pulse parameters $p_0\mapsto p_0+\delta p$ to further minimize some experimental objective characterizing gate performance. In order to reduce the number of parameters that must be calibrated in experiment, we consider the mapping $U: \mathbb{R}^M \rightarrow \mathrm{SU}(d)$ taking the $M$-dimensional vector of pulse parameters $p$ to the final $d \times d$ unitary $U(p)$ induced by the pulse parameterization and Hamiltonian model. To analyze how, according to the model, a local shift in pulse parameters by $\delta p$ from $p_0$ moves the final operation in unitary space, we consider the Jacobian $J_{\tilde{U}}(p_0)$ of $\tilde{U}$, the vectorized version of $U$, evaluated at $p_0$. This provides a local linear approximation to $\tilde{U}(p_0 + \delta p)$:
\begin{equation}
        \tilde{U}(p_0+\delta p)-\tilde{U}(p_0)=J_{\tilde{U}}(p_0)\delta p+O(\delta p^2).
        \label{jacobian expansion}
\end{equation}
We take the singular value decomposition (SVD) of $J_{\tilde{U}}(p_0)$, writing
\begin{equation}
    J_{\tilde{U}}(p_0)=W\Sigma V^\dagger=\sum_{i=1}^{r} \sigma_i w_i v_i^\top,\label{svd}
\end{equation}
where $W$ and $V$ are $d\times d$ and $M\times M$ unitaries, respectively, $\Sigma$ is a $d\times M$ rectangular diagonal matrix with singular values $\sigma_i\geq 0$ along the diagonal, and where $r$ is the rank of $J_{\tilde{U}}(p_0)$, bounded above by $d^{2}-1$. We also denote by $w_i$ and $v_i$ the column vectors of $W$ and $V$, respectively. 

Equation~\eqref{svd}  provides a geometrically motivated approach to pulse-dimensionality reduction: The singular value decomposition of $J_{\tilde{U}}(p_0)$ provides a list of orthogonal directions $v_1, \dots, v_r$ in parameter space  ordered according to the strength with which they impact the final unitary locally at $p_0$. The left singular vector $w_i$ indicates the direction in $\mathrm{SU}(d)$ along which, according to the model, displacements along $v_i$ will move the final unitary. For the single qubit case, this is visualized in Fig.~\ref{fig:dim_red}.

The dimension of $\mathrm{SU}(d)$ is $d^2-1$, and for this reason, at most $d^2 - 1$ parameters are relevant for tuning up the final unitary locally at a point $p_0$, regardless of the number of pulse parameters $M$. Furthermore, depending on factors such as the expressivity of the parameterization or the operator structure of the Hamiltonian, the rank $r$ of $J_{\tilde{U}}(p_0)$ could even be less than $d^2 - 1$. Thus, locally at $p_0$, restricting calibration to the subspace spanned by $\{v_i\}$ should not in principle sacrifice any ability to traverse the local neighbourhood of $U(p_0)$.

At this point, we note that this process depends on the model itself, which may superficially conflict with the original motivation of this work: the deviation of the actual system dynamics from those predicted by the model Hamiltonian. In practice, we treat the above reasoning as a heuristic rather than as a method for making strong quantitative predictions about experimental calibration. Using the model for pulse optimization in the first place requires the belief that, to some degree, the model captures the dynamics of the system. We take this one step further by also using the model to make an informed guess about what directions in parameter space are most useful for calibration. In effect, the core assumption behind this approach is that the model is sufficiently accurate to enable pulse design and to provide guidance for calibration, but that it lacks the precision required to eliminate the need for calibration altogether when the goal is to achieve very high gate fidelities.

In order to avoid a potential over-dependence of the reduced calibration subspace on the specific model parameters used in optimization, we perform dimensionality reduction on a collection $\{J_{\tilde{U}_{c_i}}\}$, $i=1,\dots,L$, of Jacobians, where $U_{c_i}$ denotes the final unitary obtained from a numerical simulation performed with a given realization $c_i\in\mathbb{R}^K$ of $K$ model parameters. These model parameters could include qubit frequencies, coupling strengths, and so on. We aggregate information from these simulations by computing the SVD of the $Ld^2 \times M$ matrix constructed by stacking the $L$ Jacobians vertically. The choice of $L$ can be guided by the degree of confidence in the model parameters, or by knowledge of the range they are liable to assume in a realistic experimental setting.

We remark that stacking the Jacobians is just one approach to aggregating the principal components of multiple matrices, and that other strategies could be employed instead. The approach used here is referred to as ``SVD-stack'' in \cite{baharav2025stacked}, and could be refined by weighting the matrices according to some function designed to capture some property of the model parameters, e.g., a Gaussian probability distribution centered about their expected average value. We remark on some desirable properties of this construction in idealized scenarios. If, for example, $J_{\tilde{U}_{c}}(p_0)$ is insensitive to variations in $c$, then the stacked matrix will consist of $L$ repetitions of a single matrix, and the right singular vectors reduce to the case of a single $c$ ($L=1$), while each left singular vector appears stacked $L$ times. In this scenario, there will be at most $d^2-1$ nonzero singular values. Next, consider the extreme scenario where the Jacobians $J_{\tilde{U}_{c_i}}(p_0)$ have orthogonal right singular vectors for each $c_i$ due to $\tilde{U}_c$ having a strong sensitivity to the choice of model parameters $c$. In this case, the SVD of the stacked matrix will have many nonzero singular values of a comparable weight [up to $\mathrm{min}\{L(d^2-1),M\}$], making it difficult to isolate a small set to calibrate with. In this case, the SVD serves as a witness that the ideal local calibration subspace is heavily dependent on the model parameters.

\begin{figure}
    \centering
    \includegraphics[width=0.95\linewidth]{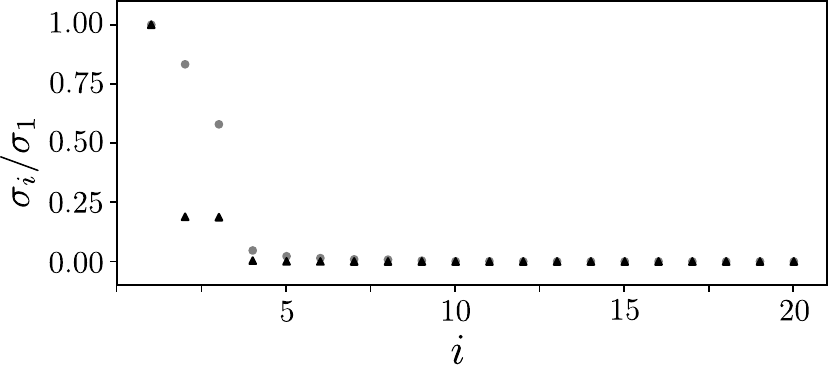}
    \caption{First 20 singular values of the stacked matrix of Jacobians, normalized by the largest singular value, computed with small fractional variations in the model (Hamiltonian) parameters for the two concrete scenarios considered in Sec.~\ref{sec:robust} (gray circles) and Sec.~\ref{sec:spectator} (black triangles).}
    \label{fig:svd-dist}
\end{figure}

In this work, the situation observed in practice lies somewhere in between the extreme scenarios discussed in the preceding paragraph: The $J_{\tilde{U}_{c_i}}(p_0)$ aren't identical, but the Jacobian is not extremely sensitive to the range of variation in model parameters deemed experimentally relevant for the system on which experimental calibration was performed. In such a scenario, the right-singular vectors of the individual Jacobians $J_{\tilde{U}_{c_i}}$ maintain significant overlap for $c_i\neq c_j$, and the nonzero singular values of the stacked Jacobian matrix therefore remain dominated in magnitude by the first $d^2-1$. The distributions of the first 20 singular values obtained for the concrete examples considered in Secs.~\ref{sec:robust} and \ref{sec:spectator} are shown in Fig.~\ref{fig:svd-dist}, ordered by decreasing size and normalized by the largest singular value. In both cases, we see that the first three singular values ($=d^2 -1$) are significantly larger than the fourth, supporting the intuition that the right-singular vectors of the Jacobians vary sufficiently slowly (as a function of the model parameters) to allow for the identification of a low-dimensional subspace of the input parameter space $\mathbb{R}^n\subset \mathbb{R}^M$ with $n\ll M$ to which pulse calibration can be restricted.

To generate the data presented in Sec.~\ref{section:pulse}, we used the first four calibration directions (spanning the calibration subspace $\mathbb{R}^4\subset \mathbb{R}^M$) and therefore performed on-device pulse calibration by tuning a vector $\bm{x}=(x_1,\dots,x_4)$ of four parameters $x_i$, which control the modification $\delta p(\bm{x})$ to the original waveform according to  
\begin{equation}\label{delta-p}
    \delta p(\bm{x})=\sum_{i=1}^4 x_i v_i,\quad \bm{x}\in\mathbb{R}^4,
\end{equation}
where $v_i\in\mathbb{R}^M$ are the first four right singular vectors of the SVD stack.

Since the qubits considered in this work are weakly anharmonic transmon qubits, the numerical simulations leading to $p_0$ were performed in some expanded Hilbert space with a cutoff dimension $d_c>2$. The matrices $U_c$ involved in the construction of the SVD stack  were taken to be the (not necessarily unitary) upper-left $2\times 2$ blocks of the $d_c\times d_c$ unitaries obtained from simulation.

\section{Pulse design and experimental results}\label{section:pulse}

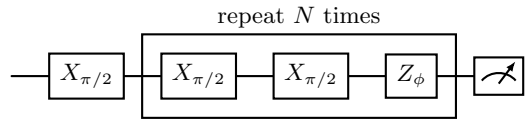
\begin{figure}[h!]
\begin{center}
\begin{quantikz}    
& \gate{X_{\pi/2}} &
\gate{X_{\pi/2}} \gategroup[1,steps=3]{repeat $N$ times} &
\gate{X_{\pi/2}} &\gate{Z_\phi}
&\meter{}
\end{quantikz}
\end{center}
\caption{General error-amplification sequence used in calibration with various values of $\phi$ and $N$. The expected outcome for an ideal $X_{\pi/2}$ gate is $\langle Z \rangle = 0$.}
\label{fig:amplification_sequences}
\end{figure}

We demonstrate the utility of this protocol towards two applications: (i) the generation of a single-qubit $X_{\pi/2}$ gate robust against variations in the pulse amplitude and detuning, and (ii) an $X_{\pi/2}$ gate robust against off-resonant spectator interactions. For both pulses, we follow the same general workflow consisting of three phases: (1) A numerical optimization to determine, according to the model, pulse parameters $p_0$ implementing the desired gate with the desired robustness properties, (2) Dimensionality reduction of the pulse parameterization at the point $p_0$, as per Eqs.~\eqref{svd} and \eqref{delta-p}, yielding the reduced parameterization $\{x_i\}$ used to calibrate the pulse in experiment. (3) Experimental calibration via tune-up of $\{x_i\}$, using the set of error-amplification experiments in Fig.~\ref{fig:amplification_sequences}.

We calibrate the pulses using the measurement statistics output by these error-amplification sequences via iterated line searches on the parameters $\{x_i\}$. Before implementing the above procedure, we also scale the pulse amplitude to align simulated and experimental results in an amplitude sweep experiment. See Appendices \ref{app:pulse_optimization} and \ref{sec:calibration} for the technical details.

\subsection{$X_{\pi/2}$ with amplitude and detuning robustness}\label{sec:robust}

System parameters like qubit frequencies and coupling strengths drift over time, requiring that gates be recalibrated periodically for optimal performance. The current IBM default pulse shape implements a Derivative Removal by Adiabatic Gate (DRAG)~\cite{motzoi2009simple} and consists of a Gaussian pulse supplemented by an out-of-phase component designed to reduce leakage to higher transmon levels. A DRAG pulse is parameterized by four numbers only: the central frequency of the pulse, its duration, the amplitude of the Gaussian, and a DRAG parameter characterizing the size of the out-of-phase component. When a DRAG pulse is applied to a qubit, the angle of rotation on the Bloch sphere is approximately linear in the product of the pulse amplitude and the strength of the qubit coupling to the driving field. Variations in the pulse amplitude or coupling strength will therefore lead to over- or under-rotation errors. Drift of the qubit frequency will lead to errors as well due to the pulse no longer being resonant with the qubit. For a pulse like DRAG whose performance degrades appreciably in the presence of drifting system parameters, maintaining good gate fidelities requires that the qubit frequency be re-characterized and the pulse amplitude re-calibrated. A common goal in pulse-shaping is therefore to design gates that are robust to small variations in both the drive amplitude (to combat drifting coupling strengths) and drive detuning (to combat drifting qubit frequencies)~\cite{carvalho2021error,kuzmanovic2024high}. Gates robust to these variations could potentially improve system stability by maintaining high gate fidelities in the presence of drifting system parameters.

When designing the pulse, we model the transmon as a weakly anharmonic oscillator with lab-frame Hamiltonian:
\begin{align}\label{transmon-hamiltonian}
\begin{aligned}
    H_0(t) &= \omega a^\dagger a +  \frac{\alpha}{2} a^\dagger a(a^\dagger a-1) +  r s(t)(a + a^\dagger),
\end{aligned}
\end{align}
where $\omega$ is the qubit frequency, $\alpha$ is the anharmonicity, $r$ is the coupling strength of the qubit to the driving field, and $s(t)=I(t)\mathrm{cos}(\omega t)+Q(t)\mathrm{sin}(\omega t)$ is a drive applied to the qubit. Variations in the qubit frequency and drive strength are modelled via the additional Hamiltonian
\begin{align}\label{perturb-hamiltonian-robust-pi/2}
\begin{aligned}
    V_c(t) &= c_0\omega a^\dagger a + c_1 r s(t)(a + a^\dagger),
\end{aligned}
\end{align}
where $c_0$ and $c_1$ are perturbative parameters representing fractional shifts in $\omega$ and $r$, respectively. Note that while the qubit parameters shift over time, typical values for the experiments described here are: $\omega / (2 \pi) = 4.74$GHz, $\alpha / (2 \pi) = -350$MHz, and $r / (2 \pi) = 85$MHz. 

\begin{figure}
    \centering
    \includegraphics[width=\linewidth]{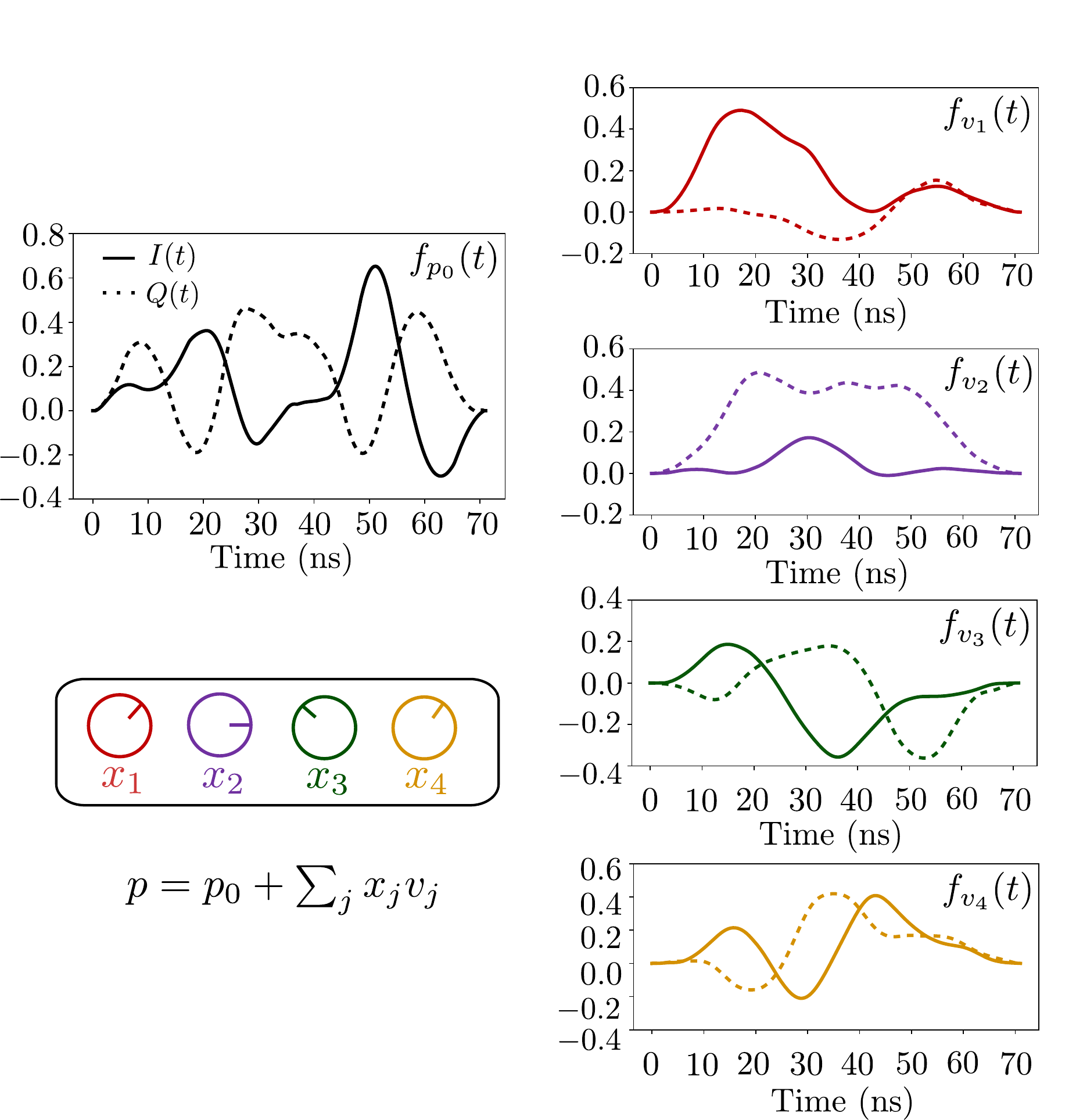}
    \caption{Waveforms $f_{p_0}(t), f_{v_i}(t)$ associated with the initial vector $p_0$ of optimized pulse parameters and the four parameter-space directions $v_1, \dots, v_4$ resulting from the dimensionality-reduction procedure, for the amplitude- and frequency-robust $X_{\pi/2}$ pulse. While the parameterization, as outlined in Appendix \ref{sec:parameterization}, contains a non-linear operation, small changes to the pulse parameters $p_0 \rightarrow p_0 + \sum_jx_jv_j$ can roughly be thought of as perturbing the envelope $f_{p_0}$ by adding the waveforms $f_{v_j}$ in various linear combinations.}
    \label{fig:pulse-envelope-and-directions}
\end{figure}

\begin{figure*}
    \centering
    \includegraphics[width=\linewidth]{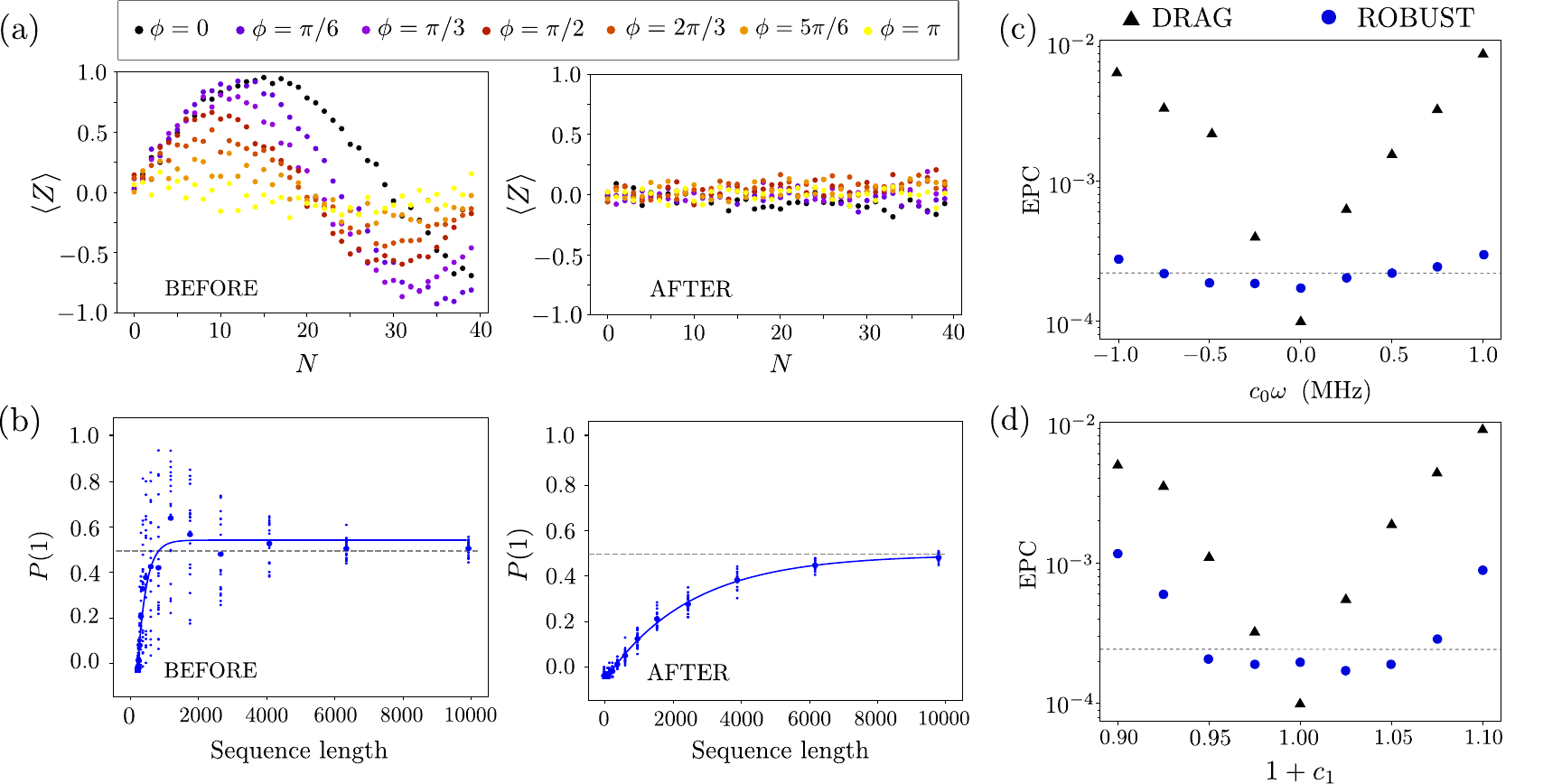}
    \caption{(a) Calibration data for several values of $\phi$ both before calibration (left) and after calibration (right). For an ideal $X_{\pi/2}$ gate, $\langle Z\rangle=0$ for all values of $\phi$ and $N$. (b) Randomized benchmarking (RB) data for the amplitude- and detuning-robust $X_{\pi/2}$ pulse both before and after calibration, where $P(1)$ is the probability of measuring the qubit in its excited state. RB was performed with a fixed gate decomposition where all Cliffords were decomposed as a product of the form $Z_{\alpha_1} X_{\pi/2}Z_{\alpha_2} X_{\pi/2}Z_{\alpha_3}$, where $Z_\alpha$ is a rotation by angle $\alpha$ about the $Z$-axis (implemented in software~\cite{mckay2017efficient}). The primary takeaway is that prior to  calibration, the gate is only roughly approximating an $X_{\pi/2}$ gate. After calibration, the estimated EPC based on an exponential  fit to the data is $1.96 \times 10^{-4}$. (c) EPC for both an $X_{\pi/2}$ gate implemented with a 32-ns DRAG pulse (black triangles) and the amplitude- and detuning-robust $X_{\pi/2}$ pulse post-calibration. EPC is given as a function of the detuning $c_0\omega$ between the qubit splitting $(1+c_0)\omega$ and the carrier frequency $\omega$ of the pulse $s(t)$ [cf.~Eq.~\eqref{perturb-hamiltonian-robust-pi/2}]. (d) EPC as a function of rescaling variations [$r\rightarrow (1+c_1)r$] in the drive amplitude, with the black triangles and blue circles defined as in (c). In both cases, the dashed line at $\mathrm{EPC}=2.5\times 10^{-4}$ is intended as a guide to the eye. }
    \label{fig:amp-freq-results}
\end{figure*}

Starting from the model Hamiltonian 
\begin{equation}
    H_c(t)=H_0(t)+V_c(t),
\end{equation}
we optimized a $71$-ns pulse designed to implement an $X_{\pi/2}$ gate  on the 27-qubit \textit{ibm-whiplash} device \footnote{For device details see \cite{sundaresan_demonstrating_2023}, which was performed on a similar device.}, with the driving-field quadratures $I(t)$ and $Q(t)$ both described by twenty free parameters, $\{a_n\}_{n=1}^{20}$ and $\{b_n\}_{n=1}^{20}$. The parameterization, described in detail in Appendix \ref{sec:parameterization}, consists of taking a linear combination of discrete Chebyshev polynomials (in which the free parameters are the coefficients), to which we apply differentiable mappings that ensure boundedness (i.e. $|I(t)|, |Q(t)| \leq 1$) and smoothness (high frequency components are removed from $I$ and $Q$). The cost function used for gate optimization includes (i) the gate infidelity, (ii) first-order perturbation-theory terms intended to minimize the impact of the  model uncertainties given in Eq.~\eqref{perturb-hamiltonian-robust-pi/2}, (iii) leakage outside the computational subspace at the final time $T=71$ ns, and (iv) leakage averaged over the entire control sequence. We set a maximum pulse amplitude based on knowledge where the electronics remained linear, and then ran optimizations for different pulse lengths until we found a length that worked for the particular objective-function. See Appendix \ref{sec:freq_amp_optimization} for a detailed breakdown of the objective-function construction.

The numerically optimized parameters $\{a_n^{\mathrm{opt}},b_n^{\mathrm{opt}}\}$ define the vector $p_0=(a_1^{\mathrm{opt}},\dots, a_{20}^{\mathrm{opt}},b_1^{\mathrm{opt}},\dots,b_{20}^{\mathrm{opt}})\in \mathbb{R}^{40}$ used as a starting point for calibration. To construct the SVD stack, we performed $L=9$ simulations to obtain $p_0$ for all combinations of qubit-frequency detunings of $-0.01 \%$, $0 \%$, and $0.01 \%$, and deviations from the ideal drive strength of $-5\%$, $0\%$, and $5\%$. Using the pulse-dimensionality reduction procedure presented above, we then identified a four-dimensional subspace in $\mathbb{R}^{40}$ to which we would restrict pulse changes during calibration [cf.~Eq.~\eqref{delta-p}].   Using this reduced parameterization, we tune the pulse to minimize errors detected by the error-amplification sequences given in Fig.~\ref{fig:amplification_sequences}. The data for the amplification sequences both before and after calibration are shown in Fig.~\ref{fig:amp-freq-results}(a). The pre-calibration data shows that the pulse does not implement an $X_{\pi/2}$ gate experimentally, despite achieving a low gate infidelity ($\approx 10^{-6}$) in simulation, and therefore requires on-device calibration. The post-calibration data, which achieves $\langle Z \rangle \approx 0$ for a large number of $\phi$ and $N$, demonstrates that the reduced parameterization [Eq.~\eqref{delta-p}] output by the dimensionality-reduction procedure was effective in tuning the experimental data.

Gate performance is quantified via randomized benchmarking (RB) using a fixed gate decomposition where every Clifford is realized as a product of the form $Z_{\alpha_1}X_{\pi/2}Z_{\alpha_2}X_{\pi/2}Z_{\alpha_3}$. Here, $Z_{\alpha}$ is a rotation by angle $\alpha$ about the $Z$-axis (implemented in software~\cite{mckay2017efficient}) and $\alpha_i$, $i=1,2,3$, are angles that can be varied to realize different gates. RB data for the gate both before and after calibration is shown in Fig.~\ref{fig:amp-freq-results}(b). This data again demonstrates that, pre-calibration, the pulse does not perform as desired, with the `BEFORE' RB data exhibiting a significant deviation from the expected exponential decay with sequence length. However, as demonstrated by the `AFTER' RB data, a significant improvement can be achieved by calibrating using only the dimensionality-reduced parameterization [Eq.~\eqref{delta-p}], resulting in an error per Clifford (EPC) of $1.96 \times 10^{-4}$. 

After calibration, we compare the performance of the optimized pulse to that achieved by a 32-ns DRAG pulse as a function of both drive-frequency detuning [Fig.~\ref{fig:amp-freq-results}(c)] and drive-amplitude rescaling [Fig.~\ref{fig:amp-freq-results}(d)]. Immediately after calibration the DRAG pulse outperforms the optimized pulse for $c_0 = c_1 = 0$, achieving an EPC of $1.1 \times 10^{-4}$, v.s.~$1.9\times10^{-4}$ for the calibrated pulse. This is to be expected based on the DRAG pulse's shorter duration. However, the DRAG-pulse performance rapidly degrades in the presence of variation in both amplitude and detuning, while the performance of the robust pulse remains stable. This demonstrates that the robustness properties engineered through pulse optimization were carried into experiment, even post-calibration. Data from two-dimensional sweeps of the EPC as a function of both $c_0$ and $c_1$ can be found in Fig.~\ref{fig:level_set_robustness} in Appendix \ref{sec:tune_up_walkthrough}, which demonstrate that the pulse is robust to simultaneous variations in both parameters.

\begin{figure*}
\includegraphics[width=\linewidth]{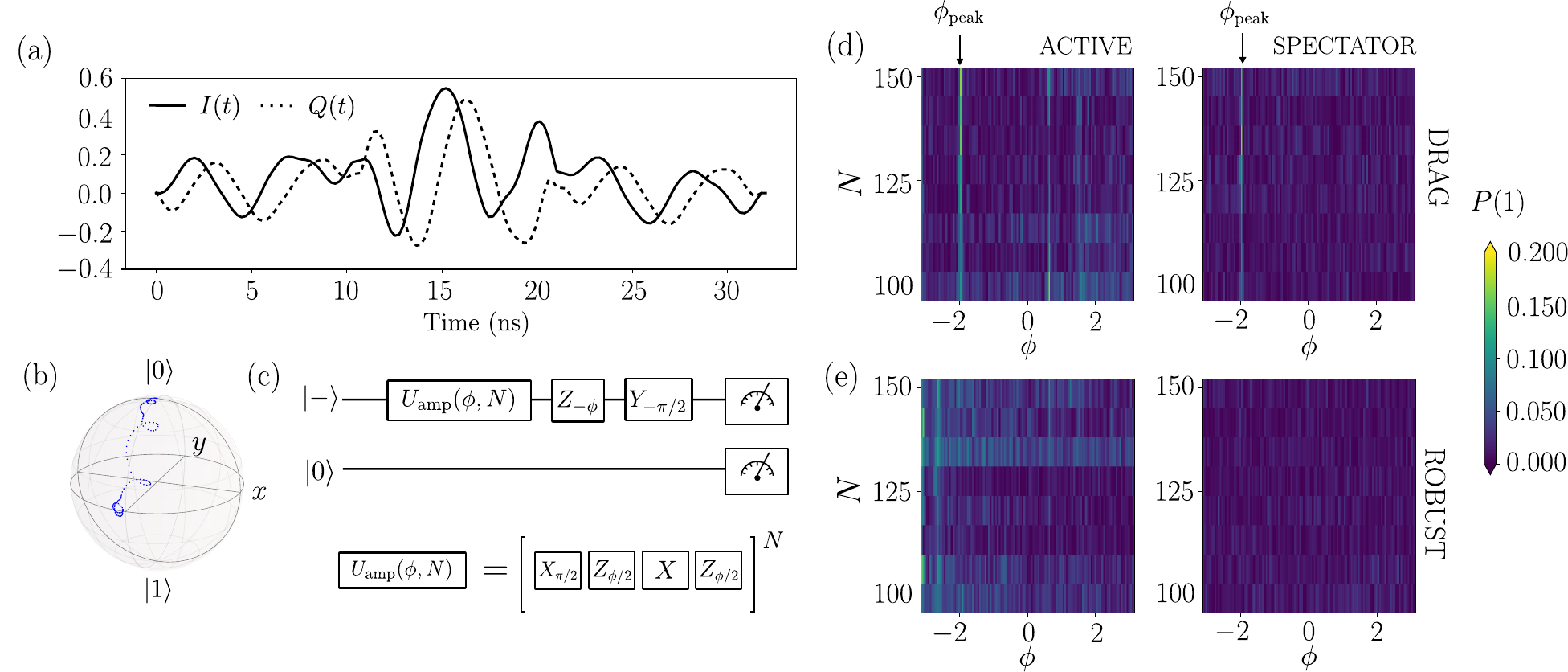}
\caption{(a) Spectator-robust waveform $s(t)=I(t)\cos{(\omega_1 t)}+Q(t)\sin{(\omega_1 t)}$ obtained from a numerical pulse optimization. (b) Bloch-sphere trajectory generated by the Hamiltonian $H(t)$ [Eq.~\eqref{eq:Hspec}] together with the waveform $s(t)$ shown in (a). (c) The phase-sweep spectroscopy data shown in (d) was obtained by preparing the active and spectator qubits in the state $\ket{-,0}$ and applying $U_{\mathrm{amp}}(\phi,N)$ for various rotation angles $\phi$ and sequence lengths $N$. This has the effect of amplifying errors due to the interaction $Z_1X_2$ with the spectator qubit for an \textit{a-priori} unknown value of $\phi\approx \phi_{\mathrm{peak}}$. (d) Phase-sweep spectroscopy data for $X_{\pi/2}$ gates implemented with 32-ns DRAG pulses, indicating correlated population inversion near $\phi_{\mathrm{peak}}\approx -2$ for the active qubit (left) and spectator qubit (right). (e) Analogous data for $X_{\pi/2}$ gates implemented with the numerically designed pulse, after closed-loop calibration on the \textit{ibm-whiplash} device. The correlated feature near $\phi_{\mathrm{peak}}\approx -2$ is significantly suppressed compared to the results achieved with DRAG.}
\label{fig:spec_fig}
\end{figure*}

\subsection{$X_{\pi/2}$ with robustness against coherent spectator-induced errors}\label{sec:spectator}

Spectator errors are a type of coherent error that occur under driving when two qubits with nearby resonant frequencies interact, often due to classical cross-talk or a fixed capacitive coupling. While spectator errors typically are not the limiting factor in today's devices, occurring with error rates $<10^{-4}$~\cite{li2023error}, they will be important to address on the road toward fault-tolerance. Furthermore, spectator interactions are a model perturbation for which pulse-based solutions within common paradigms (e.g., composite pulses) have not yet been developed.

We numerically optimized a $32$-ns pulse to generate an $X_{\pi/2}$ gate whose performance is robust against unwanted spectator interactions between neighboring qubits on the 27-qubit \textit{ibm-whiplash} device. The effective Hamiltonian describing the interaction of the driven (``active'') and spectator  qubits (having frequencies $\omega_1$ and $\omega_2$, respectively) is given by~\cite{malekakhlagh2020first}
\begin{align}\label{eq:Hspec}
    &H_c(t) = \sum_{i=1,2}\frac{\omega_i}{2}Z_i + \frac{r}{2}s(t)X_1+V_c(t),\\
    &V_c(t)=\frac{r}{2}s(t)(c_0 Z_1X_2 + c_1 X_1Z_2 + c_2 X_2)+\nu Z_1Z_2,\label{spec-perturbations}
\end{align}
where $c_{0,1}$ and $c_2$ characterize the relative strengths in units of $r$ of the cross-resonance entangling and classical cross-talk interactions, respectively, and where the term $\propto \nu$ describes an ``always-on'' $ZZ$ interaction between the two qubits.

The effects of spectator errors can be amplified and measured using phase-sweep spectroscopy~\cite{wei2024characterizing}. The procedure consists of preparing the active and spectator qubits in the state $\vert \pm,0\rangle$, applying $U_{\mathrm{amp}}(\phi,N)=(Z_{\phi/2}XZ_{\phi/2}X_{\pi/2})^N$ to the active qubit for variable rotation angles $\phi$ and sequence lengths $N$, rotating the active qubit, then measuring both qubits in the $Z$ basis [Fig.~\ref{fig:spec_fig}(c)]. The effect of the $\propto Z_1X_2$ term in Eq.~\eqref{spec-perturbations} will lead to an appreciable  probability of measuring both qubits in their excited states for $N\gg 1$ at some value of $\phi\approx\phi_{\mathrm{peak}}$~\cite{wei2024characterizing}. Since the $X$ gate in $U_{\mathrm{amp}}$ can lead to errors via the same mechanism, we implement all $X$ gates with 71-ns (lower amplitude) DRAG pulses. This reduces the impact of spectator effects due to the $X$ gates relative to those incurred during the shorter $X_{\pi/2}$ gates. 

The details of the numerical pulse design are given in Appendix \ref{sec:spectator_robust}, yielding the waveform shown in Fig.~\ref{fig:spec_fig}(a) and Bloch-sphere trajectory shown in Fig.~\ref{fig:spec_fig}(b). The pulse was calibrated using the same dimensionality-reduction procedure and calibration experiments as used for the previous pulse example. As shown in Fig.~\ref{fig:spec_fig}(d-e), the pulse eliminates the signature of spectator interactions in the phase-sweep spectroscopy data, whereas an $X_{\pi/2}$ gate implemented using a DRAG pulse of the same duration instead leads to non-negligible correlated population inversion at $\phi_{\mathrm{peak}}\approx-2$, starting near $N\approx 100$. The second bright line near $\phi\approx 0.5$ in the active-qubit data is believed to be a consequence of an interaction with a nearby two-level system, as the line disappeared after cycling the fridge. This example again demonstrates that the reduced pulse parameterization output by the dimensionality reduction procedure is sufficient to tune the gate to high fidelities, while preserving the non-trivial robustness properties built into the pulse through optimization. In RB implemented using the same fixed Clifford decomposition as in the previous example, the spectator-robust $X_{\pi/2}$ pulse achieves an EPC of roughly $7\times 10^{-5}$. This value is close to coherence limited, as implied by the decay rates $T_1 = 246.5 \;\mu s$ and $T_2 =\; 42.9 \mu s$ of the relevant qubit on the day of the experiment [see Eq.~(A3) in \cite{wei2024characterizing}].

\section{Conclusion}

In this work we introduced a method for systematically reducing the dimensionality of an arbitrary pulse parameterization, while preserving the ability to locally search the space of gates in calibration. We have demonstrated this technique in two examples and shown that, using this reduced parameterization, pulses can be tuned to achieve high gate fidelities while maintaining their various robustness properties. This approach could be used to reduce the complexity of any single-qubit-gate calibration procedure in principle, going beyond $X_\pi$ and $X_{\pi/2}$, and in future work, we will seek to automate the tune-up process to maximize its efficiency. 

The application of this method to two-qubit gates, or to other higher-dimensional systems, will likely require more careful consideration of how to choose the components of the SVD used in the dimensionality-reduction procedure. Due to the low dimensionality of single-qubit unitaries, full controllability can be achieved with a small number of parameters; in the examples presented here, we reduced the original high-dimensional pulse parameterization down to a four-dimensional effective parameterization, which one could reasonably expect an automated calibration process to handle. By contrast, the space of two-qubit unitaries is $15$-dimensional, which may already be prohibitive for direct application of black-box optimization algorithms. The dimensionality-reduction techniques explored in this work are designed to identify a subspace of the high-dimensional pulse-parameter space to which pulse variations may be restricted during on-device gate calibration. As implemented in this work, the procedure identifies a subspace with a dimension equal to that of the output space $\mathrm{SU}(d)$. It therefore provides a significant simplification to the calibration process when the initial pulse parameterization used for numerical gate design has a dimensionality far exceeding that of $\mathrm{SU}(d)$. Since the present work addresses the high dimensionality of the pulse parameterization rather than the dimensionality of the target gate, the question of how to further reduce the 15-dimensional subspace relevant to two qubit gates will require additional study. Nevertheless, we do expect this singular-value-based dimensionality reduction procedure to serve as a promising starting point for improving the on-device performance of two-qubit, numerically designed gates. More generally, it provides a much-needed interface between numerical gate design and device-based calibration, as already demonstrated here for single-qubit gates.  

\begin{acknowledgements}
We thank David C. McKay, Lev S. Bishop, and Emily Pritchett for helpful discussions and comments.
\end{acknowledgements}

\appendix
\section{Numerical pulse engineering} \label{app:pulse_optimization}
In this Appendix, we walk through the general components used for numerical pulse design. These consist of (i) the control parameterization, (ii) general approaches to constructing objective functions, and (iii) the optimization algorithm.

\subsection{Pulse Parameterization} \label{sec:parameterization}

We consider single-qubit controls of the form $s(t) = \textnormal{Re}[f(t)e^{-i \nu t}]$, where $\nu$ is a fixed carrier frequency and $f : t\in[0, T] \rightarrow \mathbb{C}$ is a complex-valued envelope. The real and imaginary part of $f(t)$ are denoted $I$ and $Q$, respectively. To perform the pulse optimization, we must choose a parameterization of the envelope $f(t)$. This parameterization must be flexible enough to explore the control space, while producing envelopes that are bounded and sufficiently smooth for realistic experimental application. When constructing our parameterizations, we allow the input parameters to be unbounded, in order to enable the use of unconstrained optimization algorithms.

As the $I$ and $Q$ components are treated independently using the same parameterization, we describe the process only for the $I$ component. Our parameterization consists of piecewise-constant envelopes with sample size $dt = 1/4.5$ ns. Let $M$ be the number of pulse parameters characterizing the $I$ component, and let $(a_1, \dots, a_M)$ denote these parameters. We construct the samples of the $I$ component as follows:
\begin{enumerate}
    \item We take a linear combination of a pre-chosen set of basis vectors representing piecewise constant envelopes $x = \sum_{i=1}^M a_i P_i$. Here, we choose the $P_i$ to be the first $M$ Chebyshev polynomials, discretized with sample size $dt$.
    \item To bound the samples, we apply the function $y = \arctan(x) / (\pi / 2)$ (applied entrywise to the samples of $x$), ensuring that the entries of $y$ lie in the interval $[-1, 1]$.
    \item We pad the vector of samples with some number of $0$s to ensure the pulse begins and ends at $0$, yielding a sample vector $z$. 
    \item Lastly, we smooth the samples by applying a Gaussian filter to $z$. The area under the Gaussian filter is $1$, ensuring that the resulting samples remain in the interval $[-1, 1]$. This convolution acts as a low-pass filter effectively eliminating high-frequency components.
\end{enumerate}
The size of the search space is set by the number of parameters $M$, the length of the pulse by the number of samples used in the discretized Chebyshev basis (together with the zero padding), and the smoothness by the parameters of the Gaussian filter. This process is guaranteed (up to numerical precision) to produce envelopes with amplitude bounded between $[-1, 1]$, and with whatever desired level of smoothness is encoded into the Gaussian filter. 

For the specific parameters used in the pulses described here, we ask the reader to refer to the optimizations in the code supplement \cite{software_supplement}.

\subsection{Robustness and time-dependent perturbation theory} \label{sec:perturbation}

To design pulses that are insensitive (``robust'') to a given perturbation to the Hamiltonian model, we use the standard perturbation-theory based approach pioneered in Average Hamiltonian Theory \cite{haeberlen_1968} and decoupling sequences \cite{mehring_1983}. In this subsection, we walk through the basic definitions and notation that we will use in the next subsection to build objective functions incorporating robustness conditions.

First, we write the model Hamiltonian as the sum of an unperturbed Hamiltonian $H_0$ and a perturbation $V_c$ characterized by a list of parameters $c=(c_1, \dots, c_k)$ corresponding to static uncertainties, 
\begin{equation}\label{model-hamiltonian}
    H_c(t)=H_0(t)+V_{c}(t),
\end{equation}
where $H_0(t)$ is independent of $c$ and $V_{c=0}(t) = 0$. We write $U_c(t) = \mathcal{T}\exp\left(-i\int_0^t ds H_c(s)\right)$, with $\mathcal{T}$ the time-ordering operator, and transform $U_c(t)$ into the toggling frame of the unperturbed Hamiltonian:
\begin{equation}
    \tilde{U}_c(t) = U_{c=0}^\dagger(t) U_c(t).
\end{equation}
For the pulses optimized in this work, we consider only first-order robustness, which depends on the first-order derivatives with respect to the model parameters of $\tilde{U}_c(t)$ about $c=0$, i.e. $\frac{\partial \tilde{U}_c(T)}{\partial c_i}\Big|_{c=0}$. Throughout this Appendix, we also work with an integral representation of the first order derivative:
\begin{equation}
    \frac{\partial \tilde{U}_c(T)}{\partial c_i}\Big|_{c=0} = -i \int_0^T dt \: U_{c=0}^\dagger(t) \left(\frac{\partial V_c(T)}{\partial c_i}\Big|_{c=0}\right)U_{c=0}(t). \label{equation:first_order_integral}
\end{equation}
These expressions are computed using functions in Qiskit Dynamics \cite{qiskit_dynamics_2023}, which are based on algorithms given in Refs.~\cite{Haas_2019,puzzuoli2023algorithms}. In particular, Ref.~\cite{puzzuoli2023algorithms} contains a full formal treatment and background of arbitrary-order perturbation-theory expansions in the multi-variable setting.

\subsection{Objective functions and optimization algorithms} \label{sec:optimization}

When  designing a control pulse, we seek to maximize the fidelity relative to the desired gate operation, maximize robustness to a set of model variations, and minimize leakage to higher-energy states of the transmon. Our optimization problem is therefore inherently multi-objective. While we ultimately scalarize the problem by taking a weighted combination of the various metrics we consider, the multi-objective nature is important for understanding some of the choices we make, as the different components of the objective can compete with each other. In this subsection, we outline at a high level the general form of the metrics used to construct the objective functions in both pulse-design examples.

First, we simulate the transmon up to some cutoff dimensions $d$. For what follows, let $A$ be the matrix mapping the computational (qubit) subspace into the fully simulated transmon space of dimension $d$, and let $P = AA^\dagger$ be the projection onto the computational subspace. Let $U(t)$ denote the unitary generated by the pulse up to time $t$, and let $T$ be the final time at which the desired gate should be implemented.

For fidelity to the target operation, let $G$ be the target unitary and $U(T)$ the simulated final operation on the transmon. To measure how well $U(T)$ implements $G$ on the qubit subspace, we use the standard fidelity metric:
\begin{equation}
    F(G, U(T)) = \frac{|\mathrm{Tr}(G^\dagger A^\dagger U(T) A)|^2}{4}. \label{fidelity}
\end{equation}
Note that here, $A^\dagger U A$ is the restriction of  $U$ to the computational  subspace. 

We use two metrics to quantify pulse-induced leakage out of the computational subspace: one which quantifies leakage at the final time, and one which quantifies leakage \emph{during} the pulse. We note that while these metrics contain some redundancy --- for example, $F(G, U(T)) = 1$ necessarily implies no leakage at the final time --- the multi-objective nature of the optimization can require adding terms to the objective function that explicitly penalize errors that are deemed particularly catastrophic. Leakage at the final time is quantified as
\begin{equation}
    L(U(T)) = \| (\id_d - P) U(T) P\|_2, \label{final_leakage}
\end{equation}
where $\|X\|_2 = \sqrt{\Tr(X^\dagger X)}$ is the Hilbert-Schmidt norm and $\id_d$ is the $d\times d$ identity matrix. The expression $L(U(T))$ quantifies how much the computational subspace is mapped into the leakage states at the final time. We also penalize leakage averaged over the duration of the pulse, with the goal of discouraging the usage of higher-energy states in the implementation of the gate. To do this, we add an ``energy penalty''-like term to the objective that penalizes occupation of states outside the computational subspace, in proportion to their energy. We do this with the following penalty term:
\begin{equation}
    \frac{1}{T}\Tr\left(\int_0^T dt\:PU^\dagger(t) \hat{N} U(t)P\right), \label{average_leakage}
\end{equation}
where $\hat{N}$ is the transmon number operator truncated at dimension $d$. 
Over the course of the pulse, excitation outside of the computational subspace is unavoidable, but the inclusion of this term prevents the optimization algorithm from finding solutions that significantly populate the higher levels. Note that the integral in Eq.~\eqref{average_leakage} takes the same form as the first-order perturbation integral in Eq.~\eqref{equation:first_order_integral}. As such, even though this term does not arise from a perturbation theory analysis, we are able to compute this expression using the time-dependent perturbation theory infrastructure built into Qiskit Dynamics.

Lastly, we consider the first-order robustness terms described in the main text.  To incorporate robustness against variations $c$ in a model parameter, we seek to minimize the impact of the first-order derivative $\frac{\partial U(T)}{\partial c}$ on the computational subspace, as quantified by the metric
\begin{equation}
    \left\|\Phi\left(\frac{\partial U(T)}{\partial c}A\right)\right\|_2, \label{no identity}
\end{equation}
where $\Phi(XA) = XA - \Tr(X) A / 2$. That is, we first consider the action of $\frac{\partial U(T)}{\partial c}$ on only the computational subspace, then remove the identity component (on the computational subspace), which acts trivially on the qubit. The norm is computed on what remains. 

For both pulse examples presented in the main text, all computations are executed using Qiskit Dynamics \cite{qiskit_dynamics_2023} and the JAX array library \cite{jax2018github}. The objective functions are linearized with weights, and the optimization problem is solved using the jaxopt library \cite{jaxopt_implicit_diff}.

\section{Pulse calibration} \label{sec:calibration}

This Appendix provides information on the strategies used for experimental pulse calibration. Pulse calibration is performed in three stages:

1. Amplitude calibration: First, we calibrate the amplitude of the pulse to ensure it is close to what was assumed in simulation. To implement the amplitude calibration,  a single application of the pulse is applied to the qubit:
    
\begin{center}
\begin{quantikz}
&\gate{X_{\pi/2}(\textrm{amp})}& \meter{} \\
\end{quantikz}
\end{center}
The above experiment is scanned over a range of amplitude-rescaling values to construct a plot of $P(1)$ versus pulse amplitude. The experimentally-generated curve is compared to a simulation-generated curve, and the overall amplitude is chosen to align the curves. This calibration also serves as a heuristic check of experimental agreement with simulation. For an $X_{\pi/2}$ pulse designed to be robust to variations in pulse amplitude, the most important feature to verify is the presence of a flat region in this curve at the chosen amplitude scaling (Fig.~\ref{fig:amp_scaling}).

2. Angle calibration (optional): 
The goal of the angle calibration is to ensure that the axis of rotation of the pulse is in the X-Y plane. This is accomplished by executing the following circuit for various values of $\theta$:\\
\begin{center}
\begin{quantikz}
&\gate{X_{\pi/2}}&\gate{Z_\theta}& \gate{X_{\pi/2}} &\meter{} \\
\end{quantikz},
\end{center}
where here, $Z_\theta=e^{-i\frac{\theta}{2}Z}$. This calibration can be understood by writing $X_{\pi/2}$ in terms of an Euler-angle decomposition of the form $Z_\alpha X_\beta Z_\gamma$, where $\beta\approx \pi/2$. By executing the circuit given above, the axis-of-rotation can be set by finding the value of $\theta$ such that the measurement outcomes are consistent with the application of an $X=X_\pi$ gate ($\theta=-\gamma-\alpha$). If this step is used, the chosen $Z_\theta$ gate is then absorbed into the definition of our $X_{\pi/2}$ gate. 

Note that this calibration step is not strictly necessary, as fine calibration (discussed next) can also be used to correct axis-of-rotation errors. (We used this procedure for the spectator-robust pulse, but not for the amplitude- and detuning-robust pulse. The spectator-robust pulse example was implemented first, and was our first attempt at this overall procedure.)
    
3. Fine calibration: We perform a fine calibration using the fact that for an ideal $X_{\pi/2}$ gate, the following experiment should return $\langle Z \rangle = 0$ for all values of the rotation angle $\phi$ independent of the number $N$ of repetitions:
\begin{center}
\begin{quantikz}    
    & \gate{X_{\pi/2}} &
    \gate{X_{\pi/2}} \gategroup[1,steps=3]{repeat $N$ times} & 
    \gate{X_{\pi/2}} &\gate{Z_\phi}
    &\meter{}
\end{quantikz}
\end{center}
The angle $\phi$ is varied is to ensure that the experiment is not tricking the user into a false positive result: If the qubit happens to be in an eigenstate of the gate, the experiment could always return $\langle Z \rangle = 0$ despite the qubit undergoing no rotation around the Bloch sphere.

This calibration makes use of the reduced parameterization obtained from the dimensionality reduction procedure described in Sec.~\ref{sec:dimensionality_reduction}. The tune-up of the (four) pulse parameters was performed via iterated one-dimensional line searches along each of the calibration directions. For each line search,  amplification-sequence data was collected for various values of the parameter, and the best-performing parameter was chosen (based on the amplification-sequence data points being close to $0$). In practice, we performed this process starting with a small number of repetitions $N$ and small set of $\phi$ values, gradually increasing both as the quality of the amplification-sequence data improved. For both pulses, we carried out this process until the amplification-sequence data was sufficiently close to $0$ for $N$ up to $40$ or $50$, and for $7$ values of $\phi$ evenly spread across $[0, \pi]$ (cf.~Fig.~\ref{fig:amp-freq-results}). 

\section{Amplitude- and frequency-robust $X_{\pi/2}$ pulse} \label{sec:freq_amp_robust}

In this Appendix, we present information specific to the design and calibration of the amplitude- and detuning-robust $X_{\pi/2}$ pulse, including a walk-through of data drawn during the calibration process.

\subsection{Numerical optimization} \label{sec:freq_amp_optimization}

We model the driven transmon via the Hamiltonian
\begin{align}
\begin{aligned}
H_0(t)&=H_s+H_d(t)\\
&=\underbrace{\omega a^{\dagger} a+\frac{\alpha}{2} a^{\dagger} a\left(a^{\dagger} a-1\right)}_{H_s}+\underbrace{r s(t)\left(a+a^{\dagger}\right)}_{H_{d}(t)},
\end{aligned}
\label{transmon hamiltonian}
\end{align}
where $\omega$ is the transmon frequency, $\alpha$ is the transmon anharmonicity, and $r$ is the strength of the coupling to the drive. We denote by $s(t)=\mathrm{Re}[f(t)e^{i\omega t}]$ the pulse applied to the transmon, where $f(t)$ is an envelope to be optimized. We parameterize $f(t)$ via $20$ parameters for both the $I$ and $Q$ components according to the parameterization described in Section \ref{sec:parameterization}. We also include an overall phase rotation between the $I$ and $Q$ components as part of the optimization, for a total of $41$ optimization parameters.

Including static uncertainties in the qubit frequency ($c_0$) and drive strength ($c_1$) in Eq.~\eqref{transmon hamiltonian}, we get
\begin{align}
\begin{aligned}
H_c(t)&=H_0(t)+V_c(t)\\
&=\omega\left(1+c_0\right) a^{\dagger} a+\frac{\alpha}{2} a^{\dagger} a\left(a^{\dagger} a-1\right)\\
&+ r\left(1+c_1\right) s(t)\left(a+a^{\dagger}\right).
\end{aligned}
\label{Total Hamiltonian}
\end{align}
In the rotating frame of the static Hamiltonian $H_s=\omega a^{\dagger} a+\frac{\alpha}{2} a^{\dagger} a\left(a^{\dagger} a-1\right)$, Eq.~\eqref{Total Hamiltonian} reads
\begin{align}
\begin{aligned}
   \widetilde{H}_c(t)&=\widetilde{H}_0(t)+\widetilde{V}_c(t)\\
   &= \omega c_0 \tilde{N}(t) + r (1 + c_1) s(t)\tilde{X}(t),
\end{aligned}
\end{align}
where $\tilde{N}(t) = e^{iH_st}a^\dagger ae^{-iH_st}$ and $\tilde{X}(t) = e^{iH_st}(a + a^\dagger)e^{-iH_st}$.

We use a linear combination of the following objectives as the cost function to be minimized: (1) the infidelity $1 - F(X_{\pi/2}, U(T))$ of the obtained unitary  relative to the desired gate operation, with $F$ defined as in Eq.~\eqref{fidelity}, (2) the norm $\left\|\Phi\left(\partial_{c_0}U(T)A\right)\right\|_2$ quantifying the gate's robustness to variations in the transmon frequency, (3) the norm $\left\|\Phi\left(\partial_{c_1}U(T)A\right)\right\|_2$ quantifying the gate's robustness to variations in the drive amplitude, (4) the time-averaged leakage out of the computational subspace [Eq.~\eqref{average_leakage}], and (5) the final leakage out of the computational subspace [Eq.~\eqref{final_leakage}]. For these individual components, we used the weightings $(1, 5, 5, 0.1, 1)$ in the optimization, producing the pulse envelope shown in Fig.~\ref{fig:pulse-envelope-and-directions}.

\subsection{Experimental tune up and additional data} \label{sec:tune_up_walkthrough}

\begin{figure}
\centering
\includegraphics[width=0.99\linewidth]{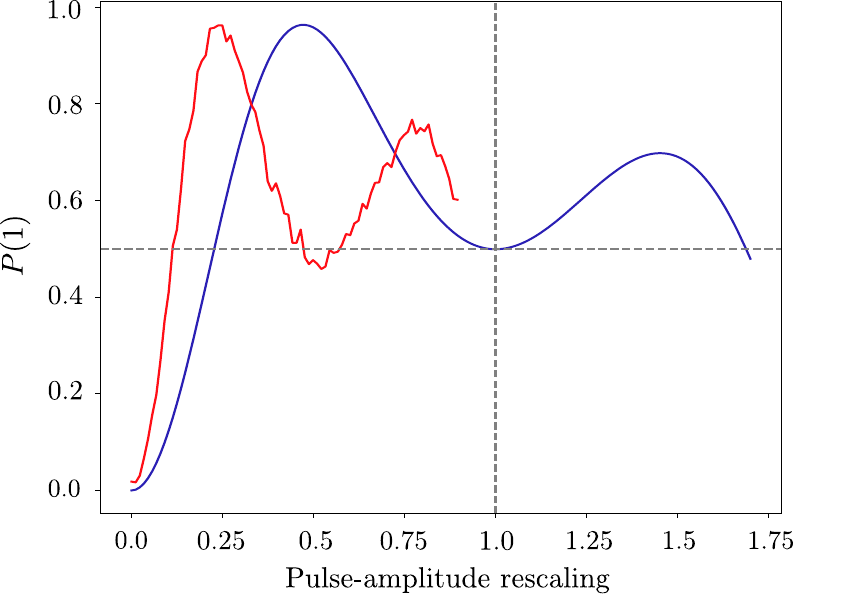}
\caption{Comparison of simulated (blue line) and experimental (red line) amplitude-calibration data. The pulse-amplitude rescaling is relative to the optimized waveform $s(t)$ found numerically. Of note is the fact that the local minimum near $P(1)=0.5$ (found in simulation by numerically solving the Schr\"odinger equation) also appears in the experimental data.  By choosing an overall rescaling of $\approx 0.5$ relative to the waveform optimized numerically, we can therefore ensure that the pulse applied in experiment is first-order insensitive to variations in pulse amplitude.  }
\label{fig:amp_scaling}
\end{figure}

We now walk through the calibration process for the amplitude- and detuning-robust $X_{\pi/2}$ pulse discussed in Sec.~\ref{sec:robust} of the main text. The amplitude-scan results are shown in Figure \ref{fig:amp_scaling}. It should be noted that the local minimum (corresponding to the robustness region where the first derivative with respect to pulse amplitude vanishes) translates from simulation into experiment. This suggests that the model used for pulse optimization captures the physics of the device relatively well, up to an overall rescaling of the pulse amplitude. Note also, however, that the experimentally measured qubit-excitation probability $P(1)$ sits visibly below $0.5$, indicating unitary error and the need for further on-device calibration. After fixing the pulse amplitude at the local minimum of the amplitude scan, we took randomized benchmarking (RB) data for the pulse, shown in Figure \ref{fig:amp-freq-results}. 

For this pulse, we proceeded directly to the fine calibration (see Appendix \ref{sec:calibration}). To calibrate the pulse, we performed sequential one-dimensional scans along each of the four calibration directions output by the dimensionality reduction procedure. Within these scans, an optimal parameter value was chosen based on how close the data came to realizing $\langle Z \rangle = 0$ for all sequence lengths $N$ and values of $\phi$. Figure \ref{fig:amp-freq-results} shows the amplification data for up to $N=40$ repetitions and seven $\phi$ values, both before and after calibration.

\begin{figure}
\centering
\includegraphics[width=0.5\textwidth]{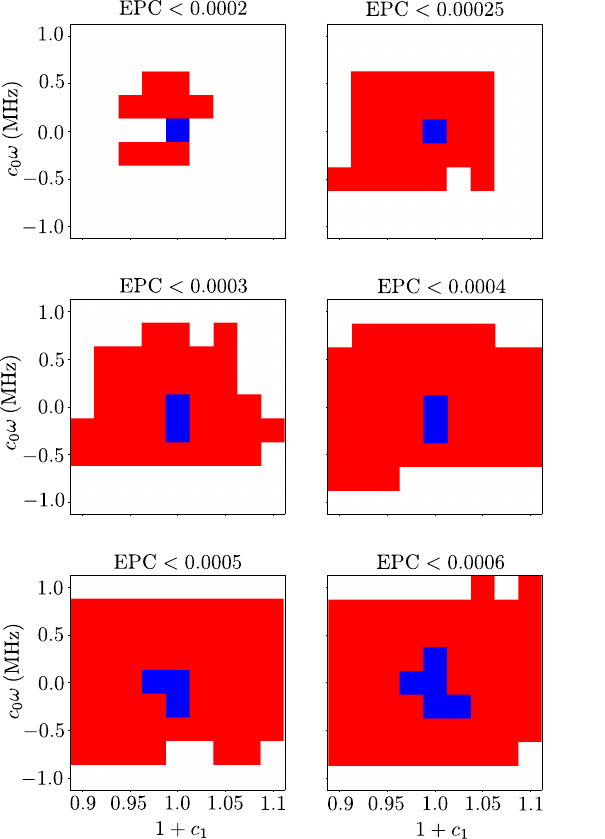}
\caption{Level sets for the error-per-Clifford (EPC) estimated by randomized benchmarking as a function of drive detuning $c_0\omega$ and pulse-amplitude rescaling $1+c_1$. Regions colored blue are those where \textit{both} the amplitude-/detuning-robust pulse and DRAG pulse achieve an EPC value below the indicated threshold, while regions colored red are those where \textit{only} the robust pulse achieves an EPC below the threshold.}
\label{fig:level_set_robustness}
\end{figure}

Figure \ref{fig:level_set_robustness} shows regions in the two-dimensional space of drive detuning and pulse-amplitude rescaling over which the robust pulse and DRAG pulse achieve an error-per-Clifford (EPC) below a given threshold, as estimated via RB with the fixed gate decomposition described in the main text. These regions demonstrate that robustness is maintained across a two-dimensional region where both the drive detuning and pulse amplitude may be simultaneously offset from their expected values.

\section{Spectator robust $X_{\pi/2}$} \label{sec:spectator_robust}

\begin{table}[h]
\begin{center}
\begin{tabular}{||c c c||} 
 \hline
 Objective & $X_{\pi/2}$ DRAG & $X_{\pi/2}$ optimized \\ [0.5ex] 
 \hline
 Gate Fidelity & 1.04e-5 & 5.72e-5 \\ 
 \hline
 Frequency Robustness & 5.92e-1 & 5.95e-1 \\
 \hline
 Amplitude Robustness & 2.79e-2 & 2.81e-2  \\
 \hline
 Cross-resonance & 3.64e-2 & 1.04e-4  \\
 \hline
Reverse cross-resonance & 3.64e-2 & 3.87e-5  \\
 \hline
 Classical cross-talk & 4.94e-6 & 4.19e-4  \\
 \hline
 Average leakage & 3.25e-4 & 6.38e-4  \\
 \hline
 Final leakage & 7.51e-4   & 4.33e-3  \\
 \hline
\end{tabular}
\caption{\label{spec_obj_table}Comparison of objective values attained for a 32-ns $X_{\pi/2}$ DRAG pulse and the 32-ns spectator-robust pulse discussed in the main text.}
\end{center}
\end{table}

In this final section, we give technical details relating to the numerical optimization of the spectator-robust pulse.
The goal is to design a pulse implementing an $X_{\pi/2}$ gate that simultaneously decouples the driven qubit from a weakly coupled spectator qubit. Following Ref.~\cite{wei2024characterizing}, we write the Hamiltonian of the driven and spectator qubits as
\begin{equation}
H_c(t) =  H_0(t) + V_c(t),
\end{equation}
where
\begin{equation}
    H_0(t) = \frac{\omega_1}{2} Z_1  + \frac{\omega_2}{2}  Z_2 + r s(t) X_1
\end{equation}
generates the ``desired'' evolution [which can be used to implement an $X_{\pi/2}$ gate on the driven qubit through proper shaping of $s(t)$], and where
\begin{align}\label{spectator-perturbation-terms}
\begin{aligned}
    V_c(t) &= r s(t) (c_0 X_1 + c_1 Z_1X_2 + c_2 X_1 Z_2 + c_3  X_2 ) \\
    &+  \nu Z_1Z_2  + c_4\frac{\omega_1}{2} Z_1  +c_5 \frac{\omega_2}{2} Z_2
\end{aligned}
\end{align}
captures the effects of system-characterization errors and ``undesired'' two-qubit interactions. When driving a single-qubit operation, the driven qubit and its spectator(s) would ideally act as totally independent systems. However, transmons with fixed-coupling are known to experience static $ZZ$ interactions~\cite{kandala2021demonstration}, and the terms $\propto c_1, c_2,c_3$ in $V_c(t)$ have been identified as producing correlated errors under single-qubit driving~\cite{wei2024characterizing}. 

In what follows, we treat $V_c(t)$ as a perturbation to the ideal Hamiltonian $H_0(t)$ and seek to design a pulse that is (ideally) first-order insensitive to $V_c(t)$. To this end, we first transform the total Hamiltonian $H(t)$ into an interaction-picture Hamiltonian $\tilde{H}(t)$ defined with respect to $(\omega_1Z_1+\omega_2Z_2)/2$, giving
\begin{align}
\begin{aligned}
\tilde{H}(t) &= r(1+c_0)s(t) \tilde{X}_1(t) \\
&+ rs(t) \left[ c_1 Z_1 \tilde{X}_2(t) + c_2 \tilde{X}_1(t)Z_2 + c_3 \tilde{X}_2(t) \right]\\
&+ \nu Z_1Z_2 + c_4 \frac{\omega_1}{2} Z_1  + c_5 \frac{\omega_2}{2} Z_2  ,
\end{aligned}
\end{align}
where
\begin{align}
&\tilde{A}_1(t) \tilde{B}_2(t)=U_0^\dagger (t) A_1B_2 U_0(t),\\
&U_0(t)=\prod_{j=1,2}e^{-\frac{i}{2}\omega_j Z_j t}.
\end{align}
In this frame, the ``desired'' term in the Hamiltonian is the first one, $rs(t) \tilde{X}_1(t)$. 

Next, we further transform the Hamiltonian $\tilde{H}(t)$ into an interaction picture defined with respect to $U(t)$, where
\begin{equation}
U(t) = \mathcal{T} \textrm{exp} \left( -i \int_0^t dt_1 rs(t_1) \tilde{X}_1(t_1) \right),
\end{equation}
yielding the following set of terms in this interaction picture:
\begin{widetext}
\begin{align}
\begin{aligned}\label{spectator-objectives}
    \dbtilde{H}(t)&=\underbrace{c_0rs(t)  U^\dagger (t) \tilde{X}_1(t) U (t)}_{\text{amplitude uncertainty}} + \underbrace{c_1rs(t) U^\dagger (t) Z_1 U (t)  \tilde{X}_2(t)}_{\text{cross-resonance term}} + \underbrace{c_2rs(t)  U^\dagger (t) \tilde{X}_1(t) U (t) Z_2}_{\text{reverse cross-resonance term}}\\&+\underbrace{c_3rs(t)  \tilde{X}_2(t)}_{\text{classical crosstalk}}
    +\underbrace{\nu U^\dagger (t) Z_1 U (t)  Z_2}_{ZZ\text{ coupling}}+\underbrace{c_4 \frac{\omega_1}{2} U^\dagger (t) Z_1 U (t) + c_5 \frac{\omega_2}{2}  Z_2}_{\text{frequency uncertainties}}.
\end{aligned}
\end{align}
\end{widetext}

In the interaction picture, the time-evolution operator is given by $\dbtilde{U}(t)=\mathcal{T}e^{-i\int_0^t dt'\:\dbtilde{H}(t')}$. In the absence of all ``undesired'' terms [$V_c(t)=0$], we would have $\dbtilde{U}(t)=\id$, so to minimize the impact of these ``undesired'' terms, we now analyze the first-order term in a perturbation expansion of $\dbtilde{U}(t)$ in powers of $\dbtilde{H}(t)$ and construct objectives to be minimized over the duration of the pulse $s(t)$. Importantly, we can write all of these objectives in a way that does not directly require simulation of the spectator qubit using the fact that
\begin{align}
    \tilde{X}_2(t) = e^{\frac{i}{2}\omega_2 t Z_2} X_2 e^{-\frac{i}{2}\omega_2 t Z_2} = \begin{pmatrix}
    0 & e^{i\omega_2 t}\\
    e^{-i\omega_2 t} & 0
    \end{pmatrix}.
\end{align}

Following the notation of Eq.~\eqref{no identity}, let $\|B\|_x = \|\Phi(B)\|_2$. That is, $\|B\|_x$ is the 2-norm of $B$ after projecting onto the set of traceless matrices (which removes the identity component, which acts trivially on the system). Furthermore, let
\begin{equation}
    (\mathcal{O}(t))_x\equiv \left\| \int_0^T dt\: \mathcal{O}(t)\right\|_x    
\end{equation}
The objectives corresponding to the various terms contributing to $\dbtilde{H}(t)$ [Eq.~\eqref{spectator-objectives}] can then be written as follows: 
\begin{enumerate}
    \item Amplitude uncertainty:
  \begin{equation*}
  \left(s(t) U^\dagger (t) \tilde{X}_1(t) U (t)\right)_x
  \end{equation*}
  \item Cross-resonance: 
  \begin{align*}
    \left(s(t)U^\dagger (t) Z_1 U (t)  \tilde{X}_2(t)\right)_x&=\left(e^{i\omega_2 t} s(t) U^\dagger (t) Z_1 U (t)\right)_x\\
    &+\left(e^{-i\omega_2 t} s(t) U^\dagger (t) Z_1 U (t)\right)_x
  \end{align*}
   \item Reverse cross-resonance:
  \begin{align*}
  \left( s(t)U^\dagger (t) \tilde{X}_1(t) U (t)  Z_2\right)_x\propto \left(s(t)U^\dagger (t) \tilde{X}_1(t) U (t)\right)_x
  \end{align*}
  \item Classical crosstalk:
  \begin{equation*}
     \left(s(t)\tilde{X}_2(t)\right)_x=\left(e^{-i\omega_2 t}s(t)\right)_x+\mathrm{c.c.}
  \end{equation*}
  \item $ZZ$ coupling: 
  \begin{equation*}
      \left(U^\dagger (t) Z_1 U (t)  Z_2\right)_x\propto \left(U^\dagger (t) Z_1 U (t)\right)_x
  \end{equation*}
  \item Frequency uncertainty:
  \begin{equation*}
    \left(U^\dagger (t) Z_1 U (t)\right)_x
  \end{equation*}
\end{enumerate}

The spectator-robust pulse discussed in the main text achieved the objective values given in Table \ref{spec_obj_table}, where we also include the objective values achieved with an $X_{\pi/2}$ DRAG pulse of the same duration. Of note are the comparable values achieved for gate fidelity, amplitude- and detuning-robustness, and gate leakage, as well as the two orders-of-magnitude improvement in the cross-resonance terms. 

\bibliography{opt_cal}

\end{document}